\documentclass[letterpaper]{emulateapj}

\begin{document}

\title[Star Formation in Disk Galaxies. III]{Star Formation in Disk Galaxies. III. Does stellar feedback result in cloud death?}

\author{Elizabeth J. Tasker\altaffilmark{1,2,3}}
\author{James Wadsley\altaffilmark{2}}
\author{Ralph Pudritz\altaffilmark{2}}
\altaffiltext{1}{CITA National Fellow}
\altaffiltext{2}{Department of Physics and Astronomy, McMaster University, 1280 Main Street West, Hamilton, Ontario L8S 4M1, Canada.} 
\altaffiltext{3}{Department of Cosmosciences, Graduate School of Science, Hokkaido University, Sapporo, Japan}

\begin{abstract}
Stellar feedback, star formation and gravitational interactions are major controlling forces in the evolution of Giant Molecular Clouds (GMCs). To explore their relative roles, we examine the properties and evolution of GMCs forming in an isolated galactic disk simulation that includes both localised thermal feedback and photoelectric heating. The results are compared with the three previous simulations in this series which consists of a model with no star formation, star formation but no form of feedback and star formation with photoelectric heating in a set with steadily increasing physical effects. We find that the addition of localised thermal feedback greatly suppresses star formation but does not destroy the surrounding GMC, giving cloud properties closely resembling the run in which no stellar physics is included. The outflows from the feedback reduce the mass of the cloud but do not destroy it, allowing the cloud to survive its stellar children. This suggests that weak thermal feedback such as the lower bound expected for supernova may play a relatively minor role in the galactic structure of quiescent Milky Way-type galaxies, compared to gravitational interactions and disk shear.
\end{abstract}

\maketitle

\section{Introduction}
\label{sec:intro}

% summary: GMC evolution is affected by SF and other forces
The Giant Molecular Clouds are the stellar nurseries of our galaxy. Forming from the coldest phase of the interstellar medium (ISM) gas, dense pockets within these extended structures collapse to birth the next generation of stars. The properties of these clouds are the controlling factor that determines the production of these gravitationally unstable dense clumps and hence the galaxy's star formation rate. Likewise, once the star is formed, its gaseous cradle will be the first environment to feel the energy it emits. The star itself is therefore a major player in the GMC's evolution, partially determining its future star-forming capabilities. As a result, this interplay between the gas dynamics and stellar feedback is of primary importance to understanding the rate of star formation in galaxy discs. 

% formation of dense clumps (and so stars) is controlled by turbulence, but what controls this is unclear
Yet, stellar feedback is not the only factor at work and exactly what does control the GMC evolution is a hotly debated subject. The balance between pressure and gravity in the cloud gas is controlled by its turbulence, which in turn can have a number of sources. Gravitational instabilities on the length of the galactic disk scaleheight inject energy that forms a turbulent cascade down to the scale of the GMCs \citep{Bournaud2010}, interactions and collisions between GMCs can shake up the gas or even cause shock waves to trigger star formation \citep{Takahira2014, Fukui2014, Tan2000} and internal to the cloud, stellar feedback from protostellar outflows, radiation, stellar winds, photoionisation and supernovae can drive turbulence that can both trigger and suppress collapse \citep{Dale2013, Lee2012, Banerjee2007, Joung2006}. 

% which turbulence generator is most important depends on cloud lifetime 
Which one of these processes dominates the evolution may depend on how long the cloud lives. For instance, in order for cloud-cloud interactions to play a significant role, the cloud must live long enough to be involved in such an event. In the first paper of this series, \citet{Tasker2009} (hereafter, Paper\,I) found that the collisions between clouds were common enough to occur roughly five times per orbital period or once every 25\,Myr at a radius of 4\,kpc. This is long compared to the free-fall time of a typical GMC which is around 4.35\,Myr for an average density of 100\,cm$^{-3}$. This means that the cloud must be somehow prevented from collapsing and converting its gas into stars if collisions are to have an impact. 

% clouds can't collapse into 100% stars but feedback may destroy clouds
A quick calculation would suggest that such a longer lifetime is extremely likely, since if the Milky Way's population of GMCs collapsed to form stars in one free-fall time, the resultant star formation rate inside the Solar circle would be approximately ${\rm M_{GMC}/t_{ff}} \approx 10^9\,{\rm M}_\odot/4.35\times 10^6\,{\rm yr} = 230$\,M$_\odot/{\rm yr}$, compared to the observed rate of $\sim 4\,{\rm M_\odot/yr}$ \citep{Krumholz2007, Williams1997, McKee1997}. However, this does not allow for the possibility the cloud is simply disrupted before it can completely convert into stars. In the scenario proposed by \citet{Murray2011}, GMCs are in a state of free-fall collapse, but the production of their first stars produces outflows that disrupt the remaining gas, resulting a total life time around $1.5-2.5\,{\rm t_{ff}}$. The conditions inside the cloud governing star formation would therefore be set principally by their properties at formation that controls the initial collapse. This view is supported by observations from \citet{Lee2012} who postulate that the expanding bubbles in star forming GMCs means that the clouds are disrupted by their internally produced feedback.

% ... or not
Yet there is also evidence that feedback has little effect on the GMCs. In simulations performed by \citet{Renaud2013}, they found that an offset developed between the densest regions of the cloud where the star originally formed and the location where the feedback occurs, due to axisymmetric drift which decoupled the newly formed star's motion from the gas. In such a case, feedback played only a minor role in the cloud's evolution and future star forming abilities. To similar effect, both \citet{Hopkins2012} and \citet{Kawamura2009} argue that a cloud can only be disrupted by feedback from a large star cluster that takes multiple free-fall times to form. As a result, the cloud can survive a finite level of stellar feedback before ultimately being dispersed by its internal processes. Based on their results from observations of the Large Magellanic Cloud (LMC), \citet{Kawamura2009} present a three-stage evolutionary sequence of the GMC's evolution. They find little difference in the physical properties of the clouds (such as radius, velocity dispersion and mass) in each of these three stages, suggesting again that local star formation activities do not impact the cloud until its final demise, which they put down to feedback from massive clusters in stage III. Even this final death does not occur instantly, and clouds can live for 10\,Myrs in stage III.

If neither feedback destroys the cloud nor they collapse entirely into stars, then the question remains as to what is controlling their evolution. Without a fresh injection of energy, turbulence within the cloud will decay over a crossing time ($t_{\rm cross} = \frac{L}{\sigma} \simeq \frac{\rm 20\,pc}{\rm 5\,km/s} = 4$\,Myr $\sim t_{\rm ff}$) and force the cloud to collapse \citep{MacLow1999}. Therefore if clouds live longer than their collapse time, additional forces must be affecting their properties, either through external interactions or non-destructive local feedback. Observations appear to favour the idea of an external driver, indicating that molecular cloud turbulence is dominated by larger-scale modes \citep{Ossenkopf2002, Heyer2004, Brunt2004, Brunt2009}. This is further supported by observations of turbulence in low star formation clouds, such as the Maddalena Cloud and the Pipe Nebular and observations of the LMC that suggest star formation is not setting the GMC properties \citep{Hughes2010}. 

In Paper I and the follow-up work in \citet{Tasker2011} (hereafter Paper II), the simulations did not include localised feedback from stars. Clouds were supported through turbulence injection from the galactic shear and cloud interactions; close encounters between which occurred far more frequently than full mergers between GMCs. The importance of such events was explored more quantitatively by \citet{Fujimoto2014}, who compared GMCs forming without feedback in the bar, spiral and outer disk regions in a simulation of M83. They found that strong interactions in the densely packed bar environment could cause a separate population of transient clouds to develop in the tidal tails of interacting Giant Molecular Associations (GMAs). Largely unbound, these clouds were not in free-fall but were at the mercy of gravitational forces from nearby objects. A similar finding that the galactic environment is a controlling force in cloud properties was made in observations by \citet{Meidt2013}, who discovered that molecular gas in M51 was prevented from forming stars by strong streaming motions that lowered the surface pressure of the cloud to prevent collapse. Further observations and simulations have noted that gas in the bar regions of disk galaxies forms stars with a lower efficiency than gas in the main disk region \citep{Fujimoto2014b, Hirota2014}.

If the majority of GMCs were actually unbound, then the free-fall time ceases to have any relevance to the cloud lifetime. Such a possibility was suggested from simulations by \citet{Dobbs2011} and observations by \citet{Heyer2001}. While pockets of dense gas may still collapse to form stars, the cloud would then not be globally falling in on itself, removing the end result of either a rapid collapse or a fast death. Other observations and simulations (including Paper I and II of this series, \citet{Benincasa2013} and \citet{McKee2007}) suggest that the clouds are borderline gravitationally bound and therefore may be encouraged to collapse, or not, by additional forces. \citet{Dobbs2013} take this further to note that the cloud and its environment cannot be entirely separated since a cloud merges and fragments over its lifetime, a process that is strongly influenced by both galactic shear and feedback. 

% this paper
In this paper, we extend the simulations performed in Paper I and Paper II to include a localised source of weak thermal energy injection. While we refer to this feedback as `supernovae', it is worth noting that it could represent any short-duration heat-depositing feedback. We explore both the evolution of clouds formed and undergoing star formation and feedback and also compare with the previous three runs presented in Papers I and II to discuss the relative importance of the effects governing GMC evolution. We find that while our feedback is enough to heavily damp the star formation within the cloud, it does not disrupt the cloud's structure. A summary of the simulations compared in this paper is shown in Table~\ref{table:runs}.

This paper is organised as follows: Section~\S 2 discusses the numerical techniques we employed, including the identification and tracking of GMCs. Section~\S 3 looks at the global properties of the galaxy disk and the interstellar medium. Section~\S4 studies the properties of the clouds themselves, both as a function of simulation time and cloud life time while section~\S5 looks specifically at the star formation rate. Section~\S6 summarises our results.

\section{Numerical Details}
\label{sec:numerics}

%%%%%%%% TABLE %%%%%%%%%%%%%%
\begin{table}[htdp]
%\scalebox{0.5}[0.5]{
\begin{tabular}{ l| p{1.6cm} p{2cm} p{1.6cm}} 
& Stars & Photoelectric heating & Local feedback \\[0.5ex] \hline \\[0.05ex]
NoSF & N & N & N  \\[0.5ex] 
SFOnly & Y & N & N \\[0.5ex] 
PEHeat & Y & Y & N \\[0.5ex] 
SNeHeat & Y & Y & Y \\[0.5ex] 
\end{tabular}
%}
\caption{Summary of the simulations compared in this paper. Run `NoSF' was published in \citet{Tasker2009} (Paper I) and runs `SFOnly' and 'PEHeat' in \citet{Tasker2011} (Paper II)}
\label{table:runs}
\end{table}
%%%%%%%% TABLE %%%%%%%%%%%%%%

\subsection{The Simulation}
\label{sec:numerics_simulation}

We performed our global simulation of the galaxy disk using {\it Enzo}; a three-dimensional adaptive mesh refinement (AMR) hydrodynamics code \citep{Enzo}. The AMR technique is particularly strong at handling multiphase fluids where many temperatures and densities co-exist in the same region, such as those found in the interstellar medium. It is also an effective technique for resolving shocks \citep{Tasker2008b}, which is a particularly important attribute when considering localised energetic feedback. {\it Enzo} has previously been used successfully to model the galaxy on this scale, including simulations that have included feedback and for our previous two papers in this series \citep[e.g.][]{Tasker2008, Tasker2009, Tasker2011}. We use a boxsize of $32$\,kpc across with a root grid of $256^3$ and an additional four levels of refinement, giving a limiting resolution (smallest cell size) of $7.8$\,pc. The location of the refined meshes was based on the \citet{Truelove1997} criteria for resolving gravitational collapse, whereby the Jeans Length must cover at least four cells. A more detailed discussion of the limits of our numerical resolution is presented in the Paper I of this series, \citet{Tasker2009}.

To evolve the gas over time, we use {\it Enzo}'s three-dimensional implementation of the {\it Zeus} hydrodynamics algorithm \citep{Stone1992} which utilizes an artificial viscosity term to handle discontinuities at shock boundaries. The variable associated with this, the quadratic artificial viscosity, was set to the default value of 2.0. We run the simulation for a total of 300\,Myr, the equivalent to just over one orbital period at the outer edge of the disk.

Radiative cooling is allowed down to $300$\,K, following the analytical cooling curve of \citet{Sarazin1987} to $T = 10^4$\,K and the extension to $300$\,K from \citet{Rosen1995}.

Star formation can potentially occur anywhere in the disk between $2.5 < r < 8.5$\,kpc (our main region for analysis outside which we do not permit stars to form), whenever a cell's density exceeds $n_{\rm H} > 100$\,${\rm cm^{-3}}$ and its temperature $T < 3000$\,K. Since $n_{\rm H} > 100$\,${\rm cm^{-3}}$ is the observed average density for a GMC, in reality star formation would occur at much higher densities within the densest part of the cloud. However, our resolution allows us to get a good handle on the GMC bulk properties, but not enough to resolve the star-forming core. We therefore allow any site within a GMC to be a potential star-forming region, but with an efficiency per free-fall time of $2$\,\%, in keeping with the observed GMC averaged star formation efficiency \citep{Krumholz2007}. We also impose a minimum star particle mass of $M_{\rm min} = 10^3$\,M$_\odot$; a numerical restriction to avoid the creation of an excessive number of star particles. In practice, our $2$\,\% efficiency means no cell ever fulfils the minimum mass criteria immediately, so the mass of `failed' star particles is tracked and a particle is formed when this quantity reaches $M_{\rm min}$. This results in a more natural, cumulative star formation process. 

We include two forms of stellar feedback in this paper. The first is the diffuse photoelectric heating included in \citet{Tasker2011}. This addition represents the ejection of electronics from dust grains via FUV photons and is proportional to the gas density, with a radial dependence described by \citet{Wolfire2003}. (See \citet{Tasker2011} for a full description). 

The second form of feedback is a thermal energy injection in the cell containing a star particle, added to the cell over its dynamical time. The equivalency of thermal and kinetic feedback has been shown by \citet{DallaVecchia2012, Durier2012} as long as the gas becomes sufficiently hot.  In this paper, we will refer to this form of feedback as `supernovae', but we note that it could represent any thermalized feedback process. Each star particle adds $10^{-5}$ of its rest-mass energy to the gas' thermal energy. This is equivalent to a supernova of $10^{51}$\,ergs for every $55$\,M$_\odot$ formed.  Since this energy is added to dense gas, the resultant cooling (including numerical and resolution effects) reduces the injected energy to 10\% of its original value, which appears predominantly as kinetic energy.  Note that this level of reduction is consistent with several studies \citep[e.g.][]{Thornton1998} who predicted that around 10\% of the SN energy should emerge as kinetic energy at late stages.  Other authors have predicted higher efficiencies, based on clustered feedback, superbubble models (e.g. \citet{MacLow1988}, 20\% kinetic and substantial thermal energy as well).  Given the ranges present in the literature, we interpret the losses in our code to be providing a lower bound on the true feedback energy in the galaxy.

\subsection{Identifying the Clouds}
\label{sec:numerics_clouds}

The galaxy is initialised as described in \citet{Tasker2009}, with a marginally stable gas disk sitting in a static background potential designed to give a flat rotation curve with circular velocity $v_c = 200$\,km/s. As the disk cools, it gravitationally fragments into objects we identify as the GMCs. 

Our analysis of the resulting density field takes place between $2.5 < r < 8.5$\,kpc in the disk, to avoid numerical effects at the disk edge and in poorly rotationally supported central region. We identify the clouds using a friends-of-friend scheme centred around peaks in the density distribution for cells with density $n_{\rm H} > 100$\,${\rm cm^{-3}}$, the average density of observed GMCs. The clouds are tracked through the simulation by comparing the population of clouds at $1$\,Myr intervals and associating clouds whose position differs from an estimated location by less than $50$\,pc. A full description of this process is presented in Paper I.

\section{Global Disk Properties}
\label{sec:global_disk}

\begin{figure*}[t]
\centering
\includegraphics[width=1.0\textwidth]{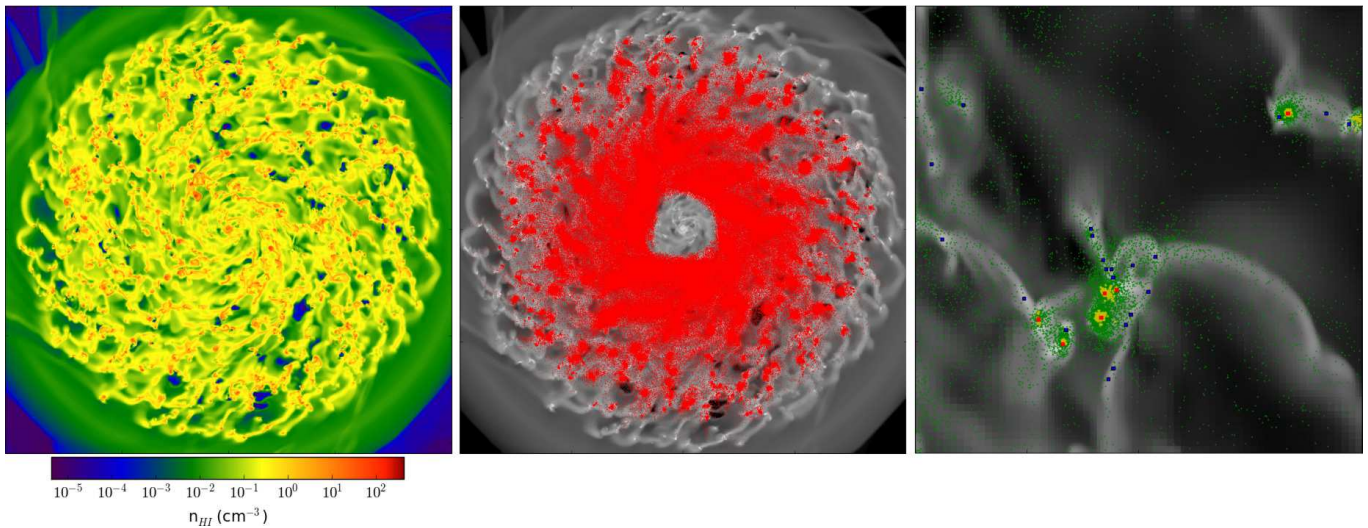}
\caption{The galactic disk at 300\,Myr. Left and center panels show 20\,kpc across the full galactic disc. The left panel shows the gas density through the mid-plane while the center image shows the gas density in grey-scale overlaid with the positions of the star particles. The right panel is a close-up of a 2\,kpc region of the ISM. Green points mark the location of all star particles while yellow points denote stars younger than 5\,Myr. Squares show the location of the identified GMCs, with pink squares denoting clouds more massive than $10^6$\,M$_\odot$.
\label{fig:diskimages}}
\end{figure*}

\begin{figure*}
\begin{center}
\includegraphics[width=\textwidth]{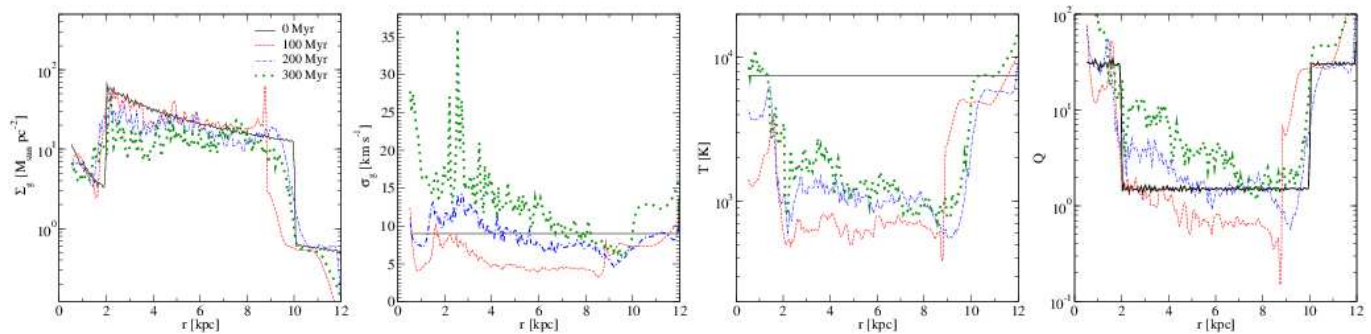}
\caption{Evolution of the azimuthally-averaged radial galactic profiles. Plots left to right show: (a) gas mass surface density ($\Sigma_g$), (b) 1D velocity dispersion of the gas ($\sigma_g$), (c) temperature ($T$) and (d) Toomre $Q$ parameter. Note, $\Sigma_g = \int_{-1{\rm kpc}}^{+1{\rm kpc}}\rho(z) d z$,  $T$ is a mass-weighted average over $-1~{\rm kpc}<z<1~{\rm kpc}$ and $Q$ makes use of $\sigma_g$ evaluated as a mass-weighted average over the same volume. 
\label{fig:diskevol}}
\end{center}
\end{figure*}

The evolution of the global structure of the disk is presented in Figures~\ref{fig:diskimages} and \ref{fig:diskevol}. Figure~\ref{fig:diskimages} is a visual view of the galaxy, with the gas surface density and stellar density shown for the entire disk and a 2\,kpc close-up surface density section shown in the third, right-hand pane. In this third image, red and blue squares mark the identified clouds' center-of-mass, with red showing the position of clouds more massive than $10^6$\,M$_\odot$. Green dots mark the location of all star particles while yellow dots denote stars younger than 5\,Myr. Figure~\ref{fig:diskevol} shows the radially averaged profiles of four disk properties; gas surface density, gas 1D velocity dispersion, gas temperature and the Toomre $Q$ measurement for disk stability. 

As with the disks presented in Paper I and II, the global gas surface density shown in the left-hand pane of Figure~\ref{fig:diskimages} shows a flocculent structure; without a time dependent potential or companion galaxy, we do not expect to excite a long-lasting grand design spiral. In Paper II, we found that the addition of photoelectric heating reduced the gravitational collapse of smaller clouds, slightly lowering the star formation rate and leading to a stronger filamentary structure in the warm ISM. This formed a nearly isobaric phase that was clearly visible as a more structured surface density image. While the dense GMCs are still connected by filaments in Figure~\ref{fig:diskimages}, they appear to form a less cohesive pattern, in closer resemblance to the no star formation run, NoSF, in Paper I. Despite the fact that photoelectric heating is included in this run, the effect of localised feedback is disrupting the filamentary structures in the warm ISM. 

This is less surprising when comparing the star particle density shown in the middle panel of Figure~\ref{fig:diskimages}. The density of stars is significantly less than either of the runs presented in Paper II, with a far sparser distribution at larger radii. In our third panel close-up in the same figure, the largest clouds marked in red are surrounded by a cluster of young (yellow) star particles, but smaller objects (blue) do not show signs of recent star formation activity. This suggests stars are predominately formed inside large clouds, which then become smaller objects with an older stellar population.

These points are quantitatively reflected in the radial profiles in Figure~\ref{fig:diskevol}. Over time, there is only a small evolution in the disk's gas surface density. Paper I showed that the static potential's flat rotation curve minimised disk evolution in the absence of star formation (an effect we desired, since we wished to compare to current star formation activity in the Milky Way, requiring minimal galactic evolution during our simulation). In contrast to this, Paper II showed a steady drop in gas surface density due to consumption of gas from star formation. At a radius of 6\,kpc, run SFOnly shows a factor of 10 decrease in gas density by 300\,Myr, compared with a drop of 2 - 3 for run SNeHeat. This is further proof that the star formation in the disk has been drastically reduced with the use of localised feedback. 

The velocity dispersion in the second panel of Figure~\ref{fig:diskevol} shows a gradual rise during the course of the simulation. This is not surprising since there is the initial fragmentation of the disk followed by cloud interactions which cause gravitational scatter. All our previous runs have shown a similar trend, however the greatest increase was seen in Paper II's SFOnly run where the velocity dispersion increased by almost a factor of 7 between 100 and 300\,Myr. This was reduced to a factor of 4 when photoelectric heating was included in run PEHeat, while in the absence of any star formation in NoSF, there was less than a factor of 3 increase. This pattern was due to SFOnly having the highest star formation rate, thereby removing the most cold gas to leave the greatest fraction of hotter, energetic gas in the disk. We see the pattern is maintained here, with the SNeHeat run showing an increase similar to that in PEHeat, although with a smaller rise in the outer parts of the disk. This again points to reduced star formation activity which allows the colder gas to remain and keep the average velocity dispersion down, but the localised feedback is visible in producing a higher average than for run NoSF.

The third panel of Figure~\ref{fig:diskevol} shows the temperature evolution of the disk. There is an increase of a factor of $3-4$ in the temperature between 100 - 300 Myr. This is small compared to disks PEHeat and SFOnly, which both showed a rise of order 30, increasing up to temperatures in excess of $10^4$\,K by 300\,Myr. This difference might seem surprising, since we are adding heat energy during the thermal feedback process, but the plot shows a mass-weighted average over the disk. This means that as cold gas is removed to form stars, the average temperature moves to that of the warm and hot ISM.  The later lower temperatures in SNeHeat are therefore once again indicative of a lower star formation rate removing cold gas. The impact of the feedback is seen if we compare with run NoSF, in the absence of any star formation. Here, the disk remained cool with temperatures close to the bottom of the radiative cooling curve at 300\,K. The star formation and localised feedback increases the radially averaged temperature to around 1000\,K, at which point it stabilises and only shows small increases with time. This is an indication the disk has reached a pseudo-steady state. (A true steady-state would not be possible without a mechanism for replenishing the gas consumed by star formation). 

The final radial profile shows the Toomre $Q$ parameter for measuring gravitational instability \citep{Toomre1964}. The disk begins borderline gravitationally stable, with $Q = 1.5$ but cools to drop below the critical $Q = 1$ and fragments. All disks in this series show a similar $Q$ radial distribution at 100\,Myr, after the initial fragmentation. After this, runs with star formation show an increase in $Q$, corresponding to the rise in the fraction of hot gas. The increase in $Q$ for SNeHeat in Figure~\ref{fig:diskevol} is slower than that for the SFOnly and PEHeat runs, again suggestive of the smaller fraction of dense gas that has been converted into stars. At 300\,Myrs, the disk still has a $Q$ value of less than 10, indicating a post-fragmentation average stability but not far from the critical value of 1.0.

\section{The Disk Interstellar Medium}
\label{sec:ism}

% PDF plot
\begin{figure}[h]
\centering
\includegraphics[width=\columnwidth]{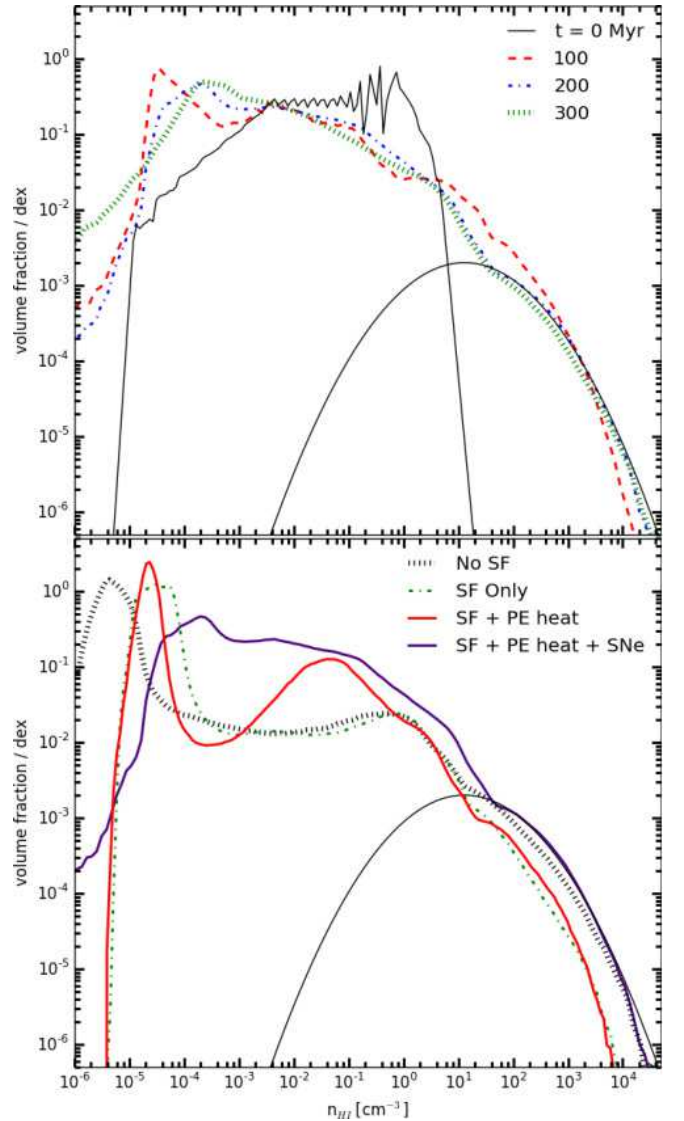}
\caption{Volume-weighted probability distribution function (PDF) for the disk over the radii $2.5 < r < 8.5$\,kpc and height -1\,kpc $< z <$ 1\,kpc. The top panel shows the evolution of the SNeHeat run during the simulation. The bottom panel shows the PDF at $t = 200$\,Myr for all four runs in this series. In both plots, the black arc marks a log-normal fit. \\
\label{fig:pdf}}
\end{figure}

The properties of the ISM can be further explored by examining the one-dimensional probability distribution functions (PDFs) in Figure~\ref{fig:pdf}. The top panel shows the volume fraction of the gas present at different densities for three different times during the SNeHeat simulation; $t = 100, 200$ and $300$\, Myr. The bottom panel shows the same result for all four runs in this series at $t = 200$\,Myr. Imposed on the PDF is a black curve showing a lognormal distribution. This curve is identical to that shown as a by-eye fit to the high density gas for the three runs in Paper I and II, with equation:

\begin{displaymath}
{\rm PDF} = \frac{1}{\sigma_{\rm PDF}\sqrt{2\pi}}e^{-\left(\ln x - \bar{\ln x}\right)^2/2\sigma^2_{\rm PDF}}
\end{displaymath}

\noindent where $x = \rho/\bar{\rho}$ and $\sigma_{\rm PDF} = 2.0$. 

In the time evolution top-panel panel for the SNeHeat run, the gas with density above 100\,cm$^{-3}$ fits a lognormal tail throughout the simulation. This develops within the first 100\,Myr, as the cold ISM fragments into clouds and generates turbulence. The profile then shows very little evolution over the subsequent 200\,Myr. Such a shape is expected for non-gravitating, isothermal turbulence \citep{Vazquez1994, Federrath2008}, but deviations have been both observed and simulated after the appearance of star-forming cores \citep{Kainulainen2009, Takahira2014}. However, the densities at which star formation is observed to occur exceed $10^4$\,cm$^{-3}$, which is beyond the point our model begins to convert gas into star particles. This prevents us from following the runaway collapse that would see such a non-lognormal extension develop over time.  

The steady-state of the high density gas is not true for all runs in this series. Looking at the lower panel in Figure~\ref{fig:pdf}, it is clear that by 200\,Myr, the dense gas in runs SFOnly and PEHeat has been reduced from the conversion into stars. By contrast, the SNeHeat run remains close to the NoSF simulation, indicative again of the lower star formation rate. Despite this off-set in high density, all runs maintain a log-normal profile for gas above 100\,cm$^{-3}$. This agrees with previous simulation results that suggest that the PDF tail is independent of the physics included in the simulation \citep[e.g.][]{Robertson2008, Tasker2008, Wada2007}.

The lower density gas below 100\,cm$^{-3}$ shows more of a dependence on the physics, with runs PEHeat and SNeHeat having a greater fraction of gas between $10^{-4} - 10$\,cm$^{-3}$. The origin of this excess is likely different in both cases. The PEHeat simulation shows a bump around 0.1\,cm$^{-3}$, corresponding to the bolstered filament structure in the disk. The SNeHeat run by contrast, has gas more evenly distributed between $10^{-4} - 1.0$\,cm$^{-3}$, possibly due to outflows from the localised feedback dispersing a section of the cloud gas and filamentary warm ISM. 

% Gas mass fraction in clouds
\begin{figure}  
\includegraphics[width=\columnwidth]{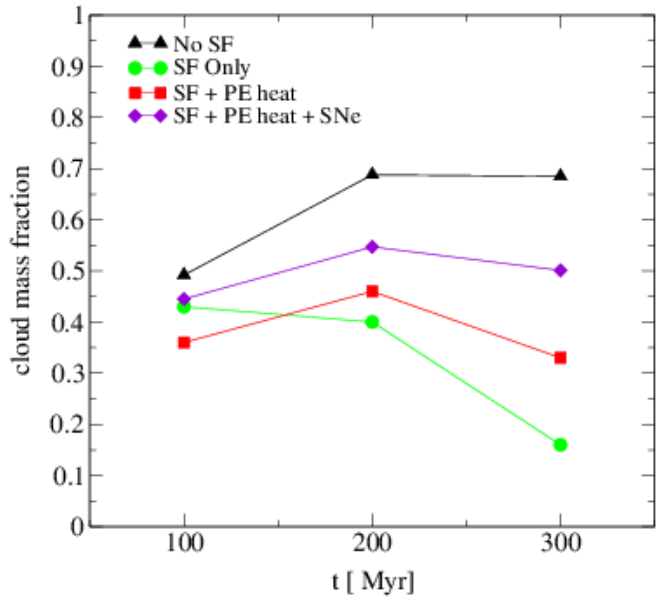}
\caption{Gas mass fraction in GMCs with $n_{\rm H} > 100$\,${\rm cm^{-3}}$ for the four runs performed in this series of papers at three different simulation times. 
\label{fig:massfrac}}
\end{figure}

The difference the physics makes to the availability of star-forming gas can be seen by comparing the mass fraction of gas in GMCs with $n_{\rm H} > 100$\,${\rm cm^{-3}}$, as shown in Figure~\ref{fig:massfrac}. Without star formation, run NoSF accumulates GMC gas for the first 200\,Myrs as the disk fragments. After this time, it flattens to a constant 0.7 fraction as gravitational interactions between clouds create a steady state from destroying and reforming the clouds in the ISM. With star formation and no feedback, SFOnly begins to consume its GMC material even before full disk fragmentation. The decrease gets steeper once the fragmentation process is complete and less new gas is collapsing into clouds. By 300\,Myr, there is only a 0.16 gas fraction in the GMCs. Adding in photoelectric heating initially leads to a lower gas fraction as smaller clouds are prevented from forming, their material held instead in the filaments. However, as larger clouds form, their gas fraction is more gradually eroded, leaving a 0.33 gas fraction in GMCs by the simulation end. In our newest SNeHeat run, the gas fraction lies close to the SFOnly run at 100\,Myr, demonstrating the disruption of the filaments from feedback seen visually in Figure~\ref{fig:diskimages}. After 200\,Myr, gas is depleted from the GMCs, but at a slower rate than either the SFOnly or PEHeat simulations. As a result, the SNe curve lies between PEHeat and the NoSF run with a 0.5 gas fraction at 300\,Myr.

% Contour plot evolution 
\begin{figure*}
\begin{center}
\includegraphics[width=\textwidth]{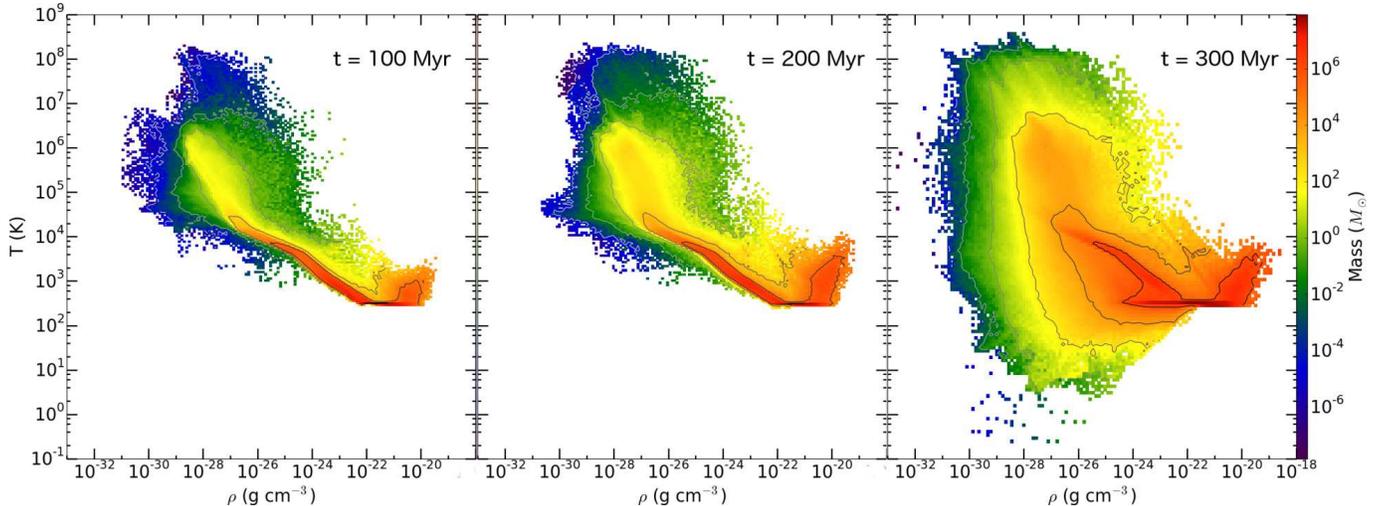}
\caption{The evolution of the mass distribution in the disk ISM in temperature versus density contour plots for run SNeHeat. The gas considered is in our main analysis region between 2.5\,kpc $< r <$ 8.5\,kpc and -1\,kpc $< z <$ 1\,kpc. 
\label{fig:contour_3times}}
\end{center}
\end{figure*}

The effect of the feedback in bolstering the lower density gas can be more clearly seen on the two-dimensional phase plots in Figure~\ref{fig:contour_3times}. This shows the mass-weighted gas distribution at the same three times at the top panel in Figure~\ref{fig:pdf} for run SNeHeat. Over time, the quantity of low density gas increases in the SNeHeat simulation, ballooning out the plot below densities $10^{-26}$\,g/cm$^3 \sim 0.01$\,cm$^{-3}$. This includes a significant quantity of adiabatic expansion below our radiative cooling limit of 300\,K. Covering a range of temperatures, this low density gas is due to the outflows from the thermal feedback, creating bubbles in the disk and pushing gas off the disk's surface where it cools. The beginning of these outflows can be seen on the right-side of each plot; dense gas above $100$\,cm$^{-3}$ collects at 300\,K inside the GMCs, but is then heated by the feedback to form high temperature regions of several 1000\,K. 

% Contour plot run comparison
\begin{figure}
\begin{center}
\includegraphics[width=\columnwidth]{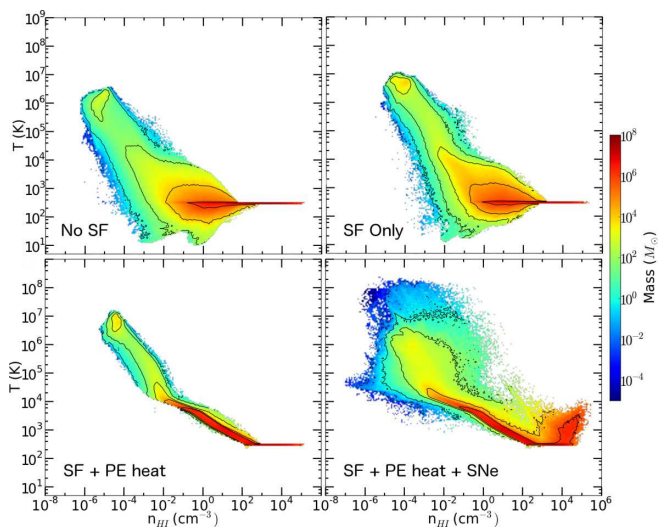}
\caption{The ISM mass distribution at $t = 200$\,Myr for the four simulations in this series. In all cases, the gas considered is in our main region between 2.5\,kpc $< r <$ 8.5\,kpc and -1\,kpc $< z <$ 1\,kpc. 
\label{fig:contour_4runs}}
\end{center}
\end{figure}

This structure looks markedly different from the previous three simulations in this series, a comparison with which is shown in Figure~\ref{fig:contour_4runs} at 200\,Myr. Without localised feedback, GMC gas remains at the radiative cooling curve minimum, producing a sharp line at 300\,K above $100$\,cm$^{-3}$. In the SNeHeat run, the densest part of the gas can be up to a factor of ten hotter as thermal energy is injected at the star formation sites. In the warm ISM between $10^3 - 10^4$\,K, the gas in the SNeHeat run resembles the PEHeat simulation, showing a higher mass in this region that either NoSF or SFOnly. This suggests that despite the lack of observed strong filaments in Figure~\ref{fig:diskimages} and \ref{fig:pdf}, the photoelectric heating continues to contribute to the contents of the warm ISM, even with the more dramatic localised feedback included. However, the reason the filaments are less marked in the SNeHeat case is also apparent, since there is a spread over six orders of magnitude in density in this region from gas ejected from the clouds. The photoelectric heating therefore continues to support gas against collapse and increase the mass in the warm ISM, but is no longer able to form an ordered filamentary structure.

\section{Cloud Evolution}
\label{sec:clouds}

% Cloud number and formation rate over time
\begin{figure*}[!t]
\begin{center}
\includegraphics[width=\textwidth]{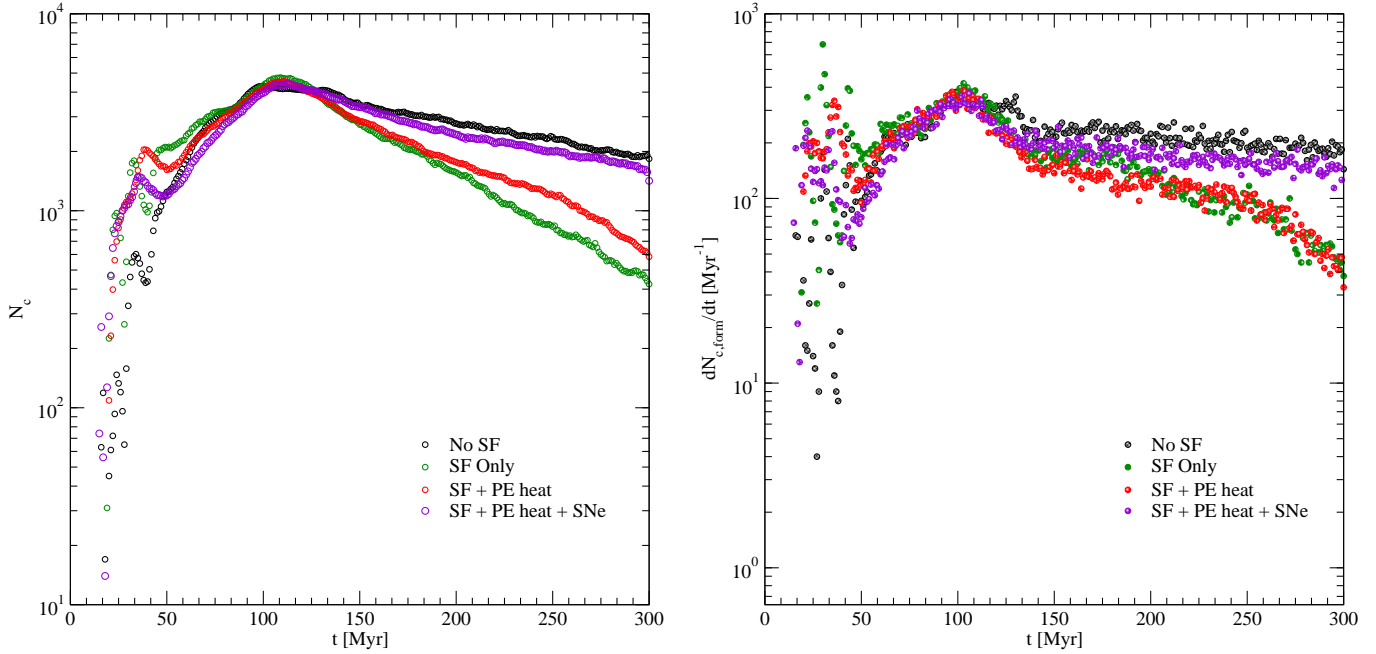}
\caption{Evolution of the cloud number for the four simulations in this series. Left-hand plot shows the total number of clouds while the right-hand plot shows their rate of formation.
\label{fig:clouds_number}}
\end{center}
\end{figure*}

We now move from a consideration of the ISM as a whole, to that of the properties of the identified clouds. GMCs are identified in the simulation as described in Section~\ref{sec:numerics_clouds}. Their number and formation rate is shown in Figure~\ref{fig:clouds_number} for each of the four simulations considered here. The left-hand panel shows the total number of clouds over the duration of the run. For all simulations, the maximum cloud count peaks at around $t = 100$\,Myr, where the disk completes its gravitational fragmentation. After that, the number of clouds decreases due to star formation, cloud mergers, dispersion due to feedback and tidal disruptions. Unsurprisingly, the highest number of clouds by 300\,Myr is for the NoSF case, since clouds here can only be destroyed in interactions with neighbouring clouds. The lowest number of clouds at the same time belongs to run SFOnly, in agreement with what we saw in Figure~\ref{fig:massfrac} where this run had the lowest fraction of cloud gas after 300\.Myr. Runs PEHeat and SNeHeat sit in-between these two extremes, with PEHeat showing a gradually increasing deviation from SFOnly over time, as the photoelectric heating slows the gas collapsing into stars. The SNeHeat run corresponds more closely to the NoSF simulation, with a small steady off-set in cloud number between $150 - 300$\,Myr. This implies that the thermal feedback not only prevents dense gas being converted into stars, but additionally does not disrupt the cloud. At earlier times around 120\,Myr, there is a small drop in cloud number in the SNeHeat run compared to the NoSF run, but then the cloud number decreases at the same rate, suggesting destruction through gravitational interactions, not feedback. The early deviation at 120\,Myr occurs when the majority of clouds are young and small, pointing to thermal feedback being able to destroy these less massive objects but failing to disrupt the main population once they have gathered mass.

The right-hand panel of Figure~\ref{fig:clouds_number} shows the rate of cloud formation. This broadly agrees with the trend seen in the left-hand plot. Runs SFOnly and PEHeat fall close together, with PEHeat showing a slightly lowered formation rate between 100 - 200\,Myr as smaller clouds remain in warm filamentary material, but a similar rate near the end of the simulation as the higher star formation rate in SFOnly reduces the gas available for clouds. The formation rate for SNeHeat run is slightly below that for NoSF due to the same photoelectric heating increasing the mass in the warm ISM as seen in Figure~\ref{fig:contour_4runs} and bolstered by the hot bubbles of gas further heating the gas surrounding the GMCs. Like the NoSF run, the formation rate remains nearly constant after 100\,Myr, indicative of the pseudo -steady state seen in the temperature profile in Figure~\ref{fig:diskevol}.

\subsection{Cloud properties}
\label{sec:clouds_properties}

% Cloud properties sim/cloud time
\begin{figure*}[P]
\begin{center}
\includegraphics[height=\textheight]{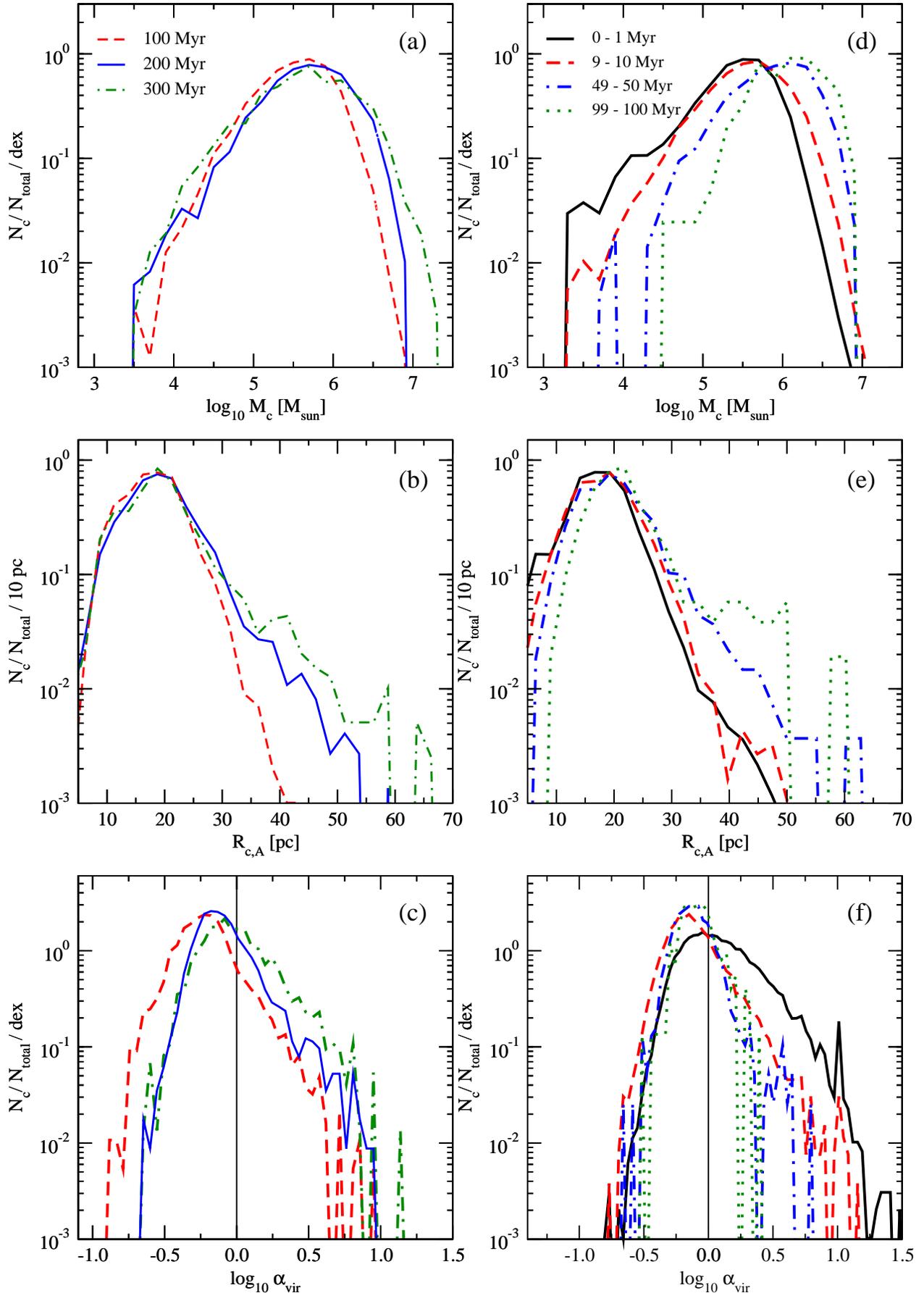}
\caption{Evolution of the cloud properties for the run including supernova feedback over simulation time (left-hand column) and cloud age (right-hand column). Plotted are cloud mass (top), cloud radius (middle) and the measure of gravitational binding, the alpha virial parameter (bottom). The clouds increase in radius over the course of the simulation and become slightly less bound. Older clouds have larger mass and a more extended structure.
\label{fig:clouds_simcloudtime}}
\end{center}
\end{figure*}

% Cloud properties simulation comparison 
\begin{figure*}[t]
%\begin{center}
\includegraphics[width=\textwidth]{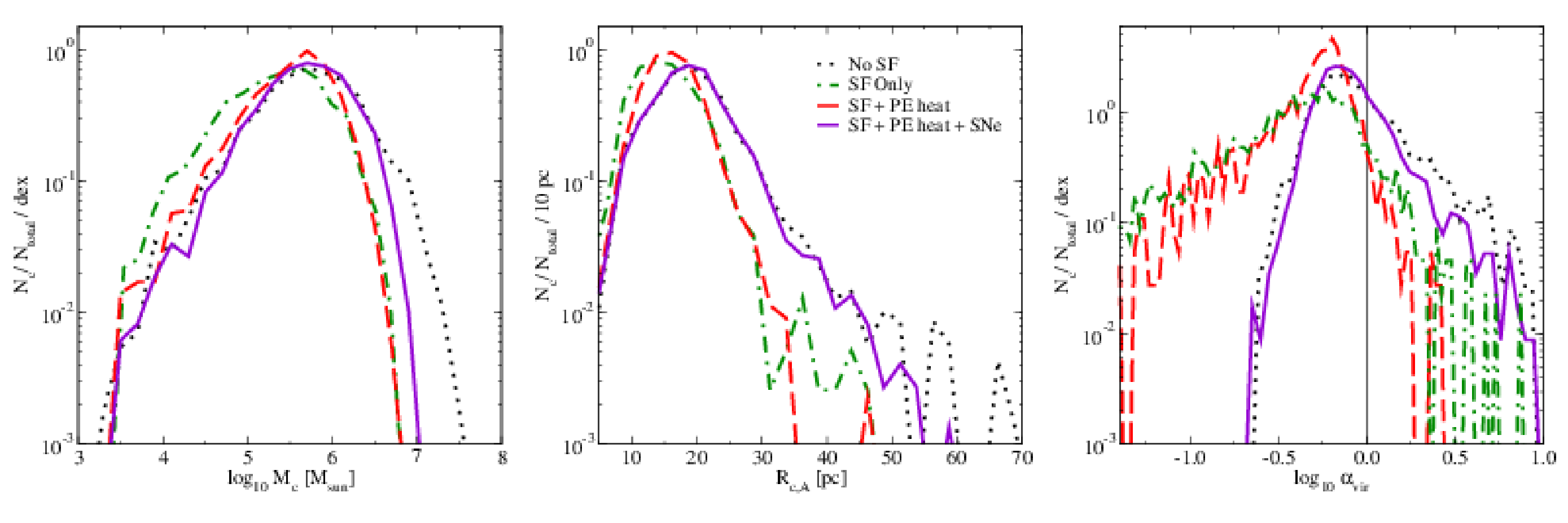}
\caption{Cloud properties at $t = 200$\,Myr for the four simulations in this series. From left to right, panels show the cloud mass, cloud radius and cloud alpha virial parameter for gravitational binding. Generally speaking, the inclusion of supernovae brings cloud properties closer to the run without any star formation.
\label{fig:clouds_comparison}}
%\end{center}
\end{figure*}

The evolution of the cloud properties over the SNeHeat simulation is shown in Figure~\ref{fig:clouds_simcloudtime}. The left-hand column of plots shows the distributions for mass (top), radius (middle) and the virial parameter estimate for gravitational binding at three simulation times, 100, 200 and 300\,Myr. The right-hand column show the same distributions for clouds of equivalent age that exist between 150 - 300\,Myr in the simulation. 

The top-left graph, Figure~\ref{fig:clouds_simcloudtime}(a), shows the evolution of the mass distribution of the cloud population. In keeping with the previous simulations in Paper I and II, the typical cloud mass at the peak of the distribution shows little change over the course of the run. Its value at $6\times 10^5$\,M$_\odot$ is independent of the physics included, with runs NoSF, SFOnly and PEHeat showing an identical typical size. This is seen clearly in the left-hand plot in Figure~\ref{fig:clouds_comparison}, which plots all four simulations at 200\,Myr for the same cloud properties. The peak value agrees well with observations of GMCs in M33, which find a peak mass of $10^5$\,M$_\odot$ \citep{Rosolowsky2003}, while the Milky Way observations by \citet{Heyer2009} report a lower typical mass of $4.8\times 10^4$\,M$_\odot$, but with the caveat that this value may be low by a factor of 3 or more \citep[mentioned by][as a private communication]{Benincasa2013}. The fact that the typical cloud mass appears to be independent of the included physics of the simulation suggests that it is likely determined by the initial gravitational fragmentation of the disk, a theory supported by simulations by \citet{Fujimoto2014}, who modelled a barred spiral galaxy and found the peak value in the cloud distributions was identical in all galactic environments (bar, spiral and outer disk). However, this is not simply the Jeans mass scale which was calculated in Paper I as $M_J \simeq 3\times 10^4$\,M$_\odot$, implying gravitational interactions between clouds must be playing a determining role. 

While the mass of the cloud population may not change significantly over time, the individual clouds do evolve. To the right of the mass distribution in Figure~\ref{fig:clouds_simcloudtime}(d), the mass distribution for clouds at different ages is plotted. As we will see later, the typical lifetime for the clouds is actually less than 20\,Myr, meaning that the distributions for the younger clouds have substantially more data. This is apparent as the peak for the whole population shown in Figure~\ref{fig:clouds_simcloudtime}(a) coincides with the peak for clouds under 50\,Myr. Clouds that do live past this are larger, more massive objects that have undergone multiple mergers to increase their mass. The minimum mass of these age distributions steadily marches higher, since smaller clouds are either consumed by their star formation and feedback or merge with bigger objects. 

Conversely, the difference in the included physics is felt in the high mass tails of the mass distribution. In the case of the SNeHeat run in Figure~\ref{fig:clouds_simcloudtime}(a), there is a small increase in the maximum mass of the total cloud population over time. This is a significantly smaller change that that seen for NoSF in Paper I, where the lack of destruction mechanisms for large clouds produced a steady increase in cloud size through mergers up to almost $10^8$\,M$_\odot$. When star formation was introduced in Paper II, SFOnly and PEHeat converted cloud mass into stars to produce a constant maximum mass of just over $6\times 10^6$\,M$_\odot$. This can be seen in the comparison in Figure~\ref{fig:clouds_comparison}(a), where the maximum cloud mass for the SNeHeat run sits in-between the runs with and without star formation. This suggests that the star formation in SNeHeat keeps the cloud mass down, but is not enough to dominate over the cloud growth from mergers. In observations, the molecular cloud mass is seen to truncate at $6 \times 10^6$\,M$_\odot$ \citep{Williams1997}. Since our clouds also include an envelope of atomic gas whose mass is considered to be between 20-100\% of the GMC \citep{Blitz1990, Fukui2009}, this comparable to an upper observational mass of $1.2\times 10^7$\,M$_\odot$.\, slightly inside what we observe in the SNeHeat run at 300\,Myr. Feasibly, the feedback is therefore suppressing the star formation slightly too strongly in this case, or other cloud disruptive forces are needed. The independence on physics that the typical mass shows in Figure~\ref{fig:clouds_comparison}(a) suggests the star formation within the clouds is lower when localised thermal feedback is included, but the cloud itself is unaffected, producing a population distribution close to that of the other runs. 

Below the mass distribution in Figure~\ref{fig:clouds_simcloudtime}(a) is the spread in cloud radii. The average cloud radius is defined at $R_{\rm c,A} = \sqrt{A_{\rm c}/\pi}$, where $A_{\rm c}$ is the projected area of the cloud in the $y -z$ plane, giving the area that would be observed from inside the galactic plane. Figure~\ref{fig:clouds_simcloudtime}(b) shows that, like the mass, the typical radius for a cloud in the SNeHeat run remains constant over time, peaking at just under 20\,pc. There is a more marked difference in the maximum cloud radius over time, with clouds at 300\,Myr being 1.5 - 2 times larger than clouds at 100\,Myr. Since this trend is not reflected in the mass, it is likely that clouds forming at late times are less concentrated. This could be due to the stronger gravitational influence of larger clouds on new clouds forming at later times or the result of the cloud undergoing thermal feedback, however two pieces of evidence suggest this is gravity. Firstly, in Figure~\ref{fig:clouds_simcloudtime}(e), the radii of different aged clouds is shown. Although significantly older clouds have a larger radius, clouds around 10\,Myr and those around 1\,Myr have similar radii. If feedback were responsible for clouds expanding, the youngest clouds should be more concentrated. A second point comes from the comparison with the previous three runs in the middle panel of Figure~\ref{fig:clouds_comparison}. The SNeHeat cloud radii distribution closely matches that of the NoSF run, although does not reach as large a maximum radius. This implies that the gravitational tug of clouds at later times --which acts in both cases-- produces a more diffuse population. For clouds with star formation but no thermal feedback, the maximum radius is reduced as gas is converted into stars. There is a small difference in the typical radius between runs SFOnly and PEHeat and the other two runs, suggesting that the clouds are more concentrated, likely due to a higher fraction of stars. 

The final two plots on the bottom row of Figure~\ref{fig:clouds_simcloudtime} compare the alpha virial parameter. This property is defined via $\alpha_{\rm vir} = 5\sigma_{\rm c}^2 R_{\rm c,A}/ (GM_{\rm {c,s}})$, where $M_{\rm c,s}$ is the combined mass of gas and stars within a cloud and $\sigma_{\rm c}$ is the mass averaged velocity dispersion of the cloud, $\sigma_{\rm c} \equiv (c_{\rm s}^2 + \sigma_{\rm nt,c}^2)^{1/2}$, with $\sigma_{\rm nt, c}$ the one-dimensional rms velocity dispersion about the cloud’s center-of-mass velocity and $c_{\rm s}$, the speed of sound. A spherical, uniform cloud with a virial parameter less than 1 is virialized and dominated by gravity.

Over the course of the simulation, the typical cloud value is slightly less than $\alpha_{\rm vir} = 1.0$, suggesting the clouds are mainly borderline virialised. In keeping with the extended radii at later times, the cloud population at 300\,Myr has a larger number of unbound objects with $\alpha_{\rm vir} > 1$. The evolution of the individual clouds in  Figure~\ref{fig:clouds_simcloudtime}(f) however, shows the opposite trend. While the typical cloud value remains borderline virialised, more clouds are unbound when younger than 1\,Myr than those that live to long ages. This is due to the effect of star formation, as gas is converted into stars, increasing the total (gas and stellar) cloud mass without expanding the radius. In the comparison between all runs in Figure~\ref{fig:clouds_comparison}(d), the SNeHeat run cloud population once again resembles the NoSF run most closely, although it has a slightly smaller population of unbound clouds, likely due to its star formation. Where there is star formation but no feedback, the clouds quickly become tightly bound, giving a similar typical value but a more extended low-virial tail. This corresponds to the lower typical radius seen in the middle panel of the same figure.

% Cloud theta comparison
\begin{figure}[!t]
\begin{center}
\includegraphics[width=\columnwidth]{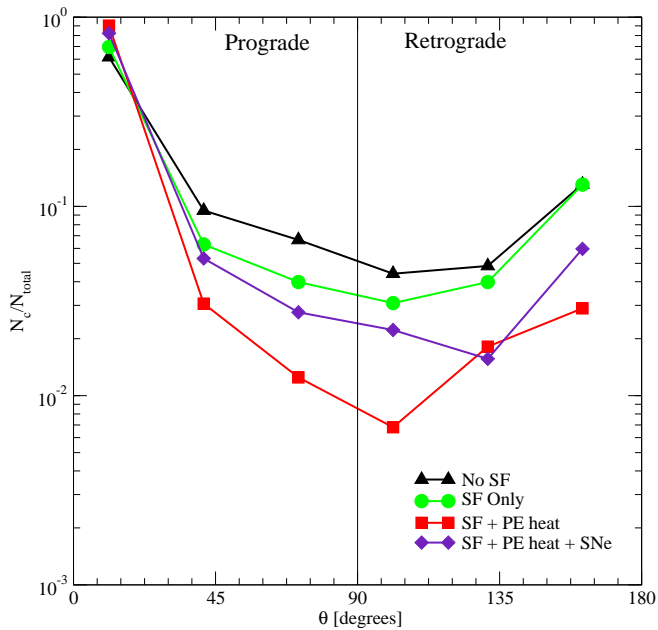}
\caption{The angle between the cloud's angular momentum axis and that of the galaxy for each of the four simulations in this series at 200\,Myr. Angles greater than $90^\circ$ denote the retrograde population of clouds that have been formed via interactions with other clouds.
\label{fig:cloudtheta_comparison}}
\end{center}
\end{figure}

An additional cloud property that can be compared is how the cloud's rotation on its own axis compares to that of the galaxy. Figure~\ref{fig:cloudtheta_comparison} shows the angular difference in the cloud and galaxy's angular momentum vector for all four runs at 200\,Myr. The majority of the clouds rotate prograde, moving in the same sense at the galaxy. This fits with the discussion in Paper I and II, where clouds are born predominantly prograde but later can be either turned via cloud interactions or born close to a large cloud which affects their rotation. This process happens in all the runs, although at 200\,Myr, the fraction of retrograde clouds varies. The NoSF run with no star formation or feedback has the highest number of retrograde clouds. This is a reflection on the cloud size, with the bigger clouds present in this simulation exerting a stronger gravitational pull that can more easily reverse the direction of smaller, nearby objects. The smallest retrograde population is from the PEHeat run, since its filamentary ISM discourages counter-rotation, as described in more detail in Paper II. The SFOnly has roughly the same number of retrograde rotators as NoSF while the SNeHeat run sits in-between NoSF and PEHeat. This is contrary to other cloud properties for SNeHeat, which more closely mimicked the NoSF simulation. The lower retrograde count is due to the inclusion of photoelectric heating which, while not marking the filaments as strongly as in PEHeat due to the outflows from the thermal feedback, still leaves an imprint on the ISM as seen in Figure~\ref{fig:contour_4runs}.

% Cloud age and mass comparison
\begin{figure}
\begin{center}
\includegraphics[width=\columnwidth]{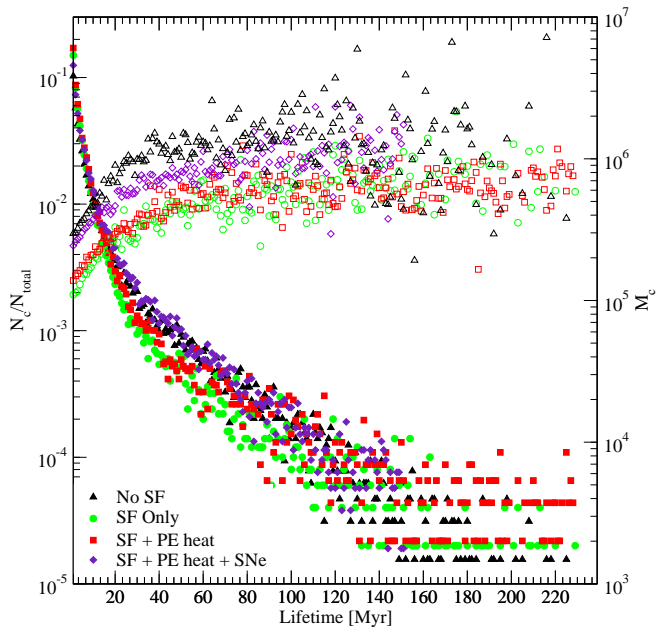}
\caption{A comparison of cloud lifetime for each of the four simulations in this series. Left-axis plots the number of clouds as a function of their lifetime while the right-axis shows the average mass the cloud has during its lifetime. Only 1\% of clouds survive beyond 20\,Myr, independent of the physics included. During this time, the average cloud mass it tightly dependent on cloud age, but later is independent.
\label{fig:cloudagemass_comparison}}
\end{center}
\end{figure}

While Figure~\ref{fig:clouds_simcloudtime} showed the distribution of clouds with very long lifetimes, the majority of clouds exist for a shorter duration. Figure~\ref{fig:cloudagemass_comparison} plots the number of clouds as a function of cloud age, showing a steep drop-off in which only 1\,\% of clouds survive beyond 20\,Myr. This is true for all runs, although for the remaining 1\,\% that do persist beyond 20\,Myr, the simulations begin to separate. 

With only star formation included, run SFOnly has the smallest population of long lived clouds. Even with no external forces operating, cloud mass will be steadily removed by the star formation. With photoelectric heating offering small support against collapse and stellar conversion, run PEHeat has a slightly larger fraction of clouds with long lifetimes. This is even more marked where supernovae feedback is included, with run SNeHeat and NoSF being once again hard to differentiate until 150\,Myr, after which only a small handful of clouds remain for the non-localised feedback cases.

Figure~\ref{fig:cloudagemass_comparison} also plots the cloud mass as a function of age on the right-hand axis. Like Figure~\ref{fig:clouds_simcloudtime}(a) and (d), this shows that the clouds that have very extended lives also have the highest mass, built up through successive mergers. Runs with the lowest star formation, either through non-inclusion (NoSF) or suppression (SNeHeat) build up the largest mass, since less is subtracted to create star particles. The mass increases steadily during the first 20\,Myr of the cloud lifetime, but then flattens to produce a maximum typically between $4\times10^5 - 3\times 10^6$\,M$_\odot$. This suggests a quasi-equilibrium between gas increase through mergers and decrease from tidal shredding and star formation. Unsurprisingly, runs SFOnly and PEheat have the smallest mass, since their star formation rate is the highest. Notably, even young clouds in these runs are born with a lower mass, reflecting the smaller quantity of gas present in the disc when the cloud analysis begins at 150\,Myr.

% Cloud merger rate simulation comparison
\begin{figure}
\begin{center}
\includegraphics[width=\columnwidth]{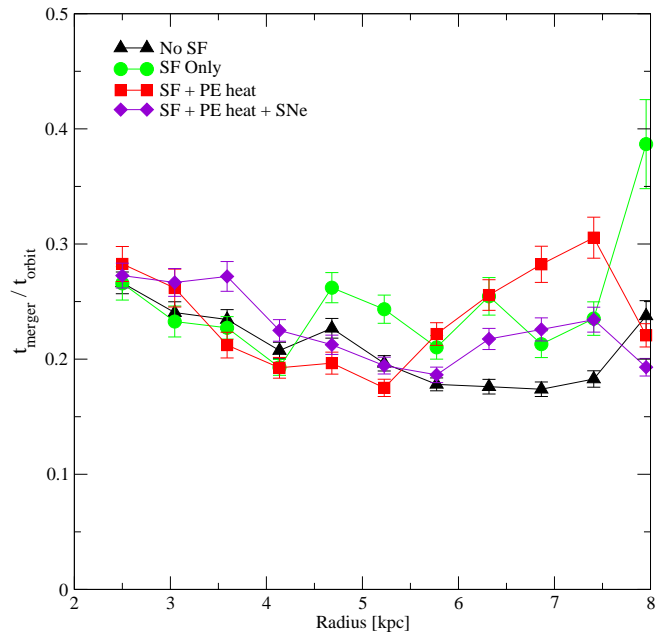}
\caption{Time between cloud mergers as a fraction of the orbital time for the four simulations in this series, averaged between t = 175 - 225\,Myr. Despite the different physics included in each run, the average merger time remains similar in all simulations.
\label{fig:merger_comparison}}
\end{center}
\end{figure}

The actual merger rate of the clouds is plotted as a fraction of the orbital period in Figure~\ref{fig:merger_comparison}. This is actually independent of both the physics included and the radius in the disk, with clouds typically undergoing a merger every $0.2-0.3$ of an orbit around the galaxy. This frequency was proposed by \citet{Tan2000} to be sufficient to explain the Kennicutt-Schmidt relation between the gas surface density and the surface density of the star formation rate \citep{Kennicutt1998}; a non-intuitive connection, since star formation happens at high volume density on small scales, which does not have an obviously straight-forward correlation with the averaged surface density. By proposing star formation could be triggered through cloud interactions, \citet{Tan2000} linked the large-scale disk properties with localised star production. Note that even though the merger rate is similar between runs, runs NoSF and SNeHeat can gain larger clouds through mergers due to the clouds living longer to undergo multiple interactions. While we have already seen substantial evidence that feedback is playing a significant role in the star formation of run SNeHeat, the frequency of the cloud collisions in all runs  indicates that interactions always play a substantial role in shaping the cloud evolution.

\section{Star Formation}
\label{sec:stars}

So far, we have speculated on the relative star formation of the runs by examining the evolution of the cloud mass. However, a more direct measure is to look at the production of the star particles themselves. 

% Cloud stellar fraction: evolution and comparison
\begin{figure*}
\begin{center}
\includegraphics[width=\textwidth]{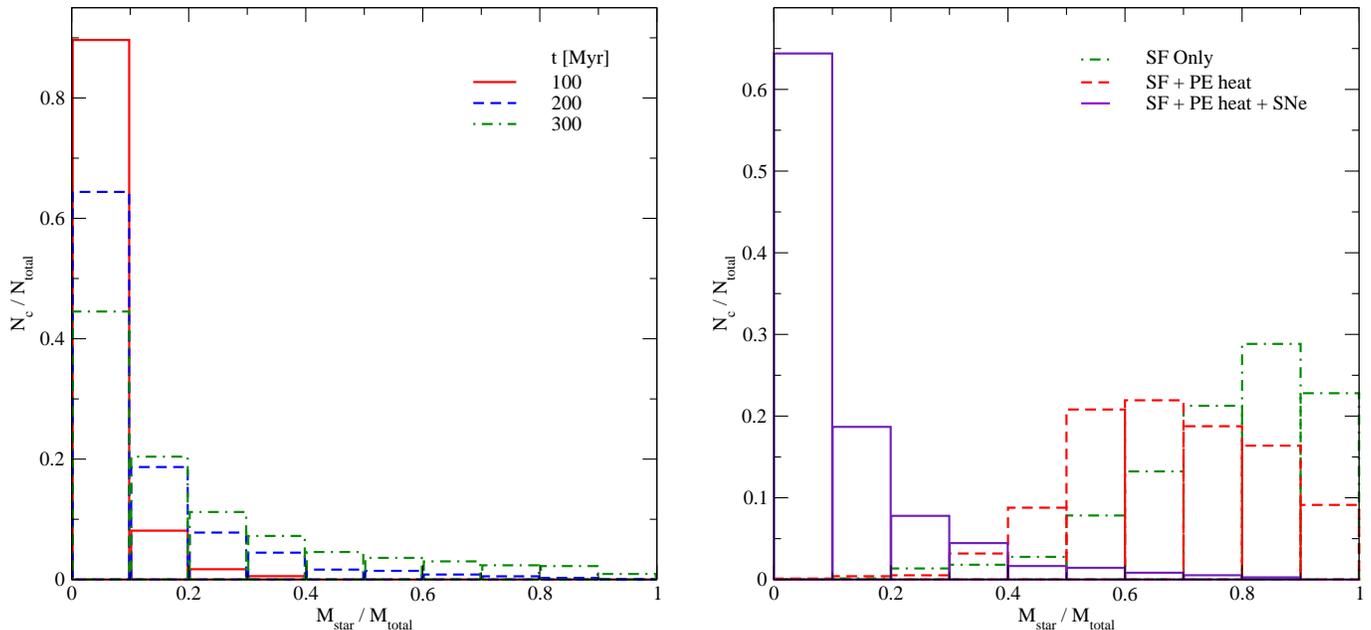}
\caption{Time between cloud mergers as a fraction of the orbital time for the simulations in this series that include active star formation, averaged between t = 175 - 225\,Myr. Despite the different physics included in each run, the average merger time remains similar in all simulations.
\label{fig:starfrac}}
\end{center}
\end{figure*}

Figure~\ref{fig:starfrac} plots the distribution of the fraction of a cloud's mass that is in stars. The left-hand panel shows the evolution of the cloud stellar fraction for the SNeHeat run, plotted at the usual three simulation analysis times of 100, 200 and 300\,Myr. The right-hand panel shows the comparison between the three star-forming runs in this series at 200\,Myr. The conditions for star particle formation are described in Section~\ref{sec:numerics_simulation} and the particles are allocated to a particular cloud if their position is within the cloud boundary or they are otherwise unallocated but within two average radii of the cloud. This second condition is necessary to allow for the decoupling of the gas and star particle motion through the disk.

On the left-hand panel of the evolution in run SNeHeat, we see a steady decrease in the fraction of high-percentage gas clouds and a corresponding increase in the fraction of clouds that are stellar dominated. However, by the end of the simulation at 300\,Myr, just under 50\,\% of clouds still have over 90\,\% of their mass as gas, while only a few percent have the majority of their mass in stars. This is a far bigger split than is found in the runs SFOnly and PEHeat where localised feedback was absent, as shown in the right-hand panel (note, different y-axis scale to display these differences more clearly). At 200\,Myr, run SFOnly has converted the greatest percentage of its clouds' gas into stars, with the highest fraction of clouds having between 80 - 90\,\% of their mass as star particles. With photoelectric heating in run PEHEeat, this fraction is reduced slightly so the most typical cloud has between 60 - 70\,\% of its mass in stars. By contrast, run SNeHeat still has an overwhelming majority of clouds predominantly gas, with 70\,\% of its clouds with less than 10\,\% star particles. This agrees with the cloud properties discussed in the previous section, where run SNeHeat showed far less sign of cloud evolution through the production of stars.

% SFR history comparison
\begin{figure}
\begin{center}
\includegraphics[width=\columnwidth]{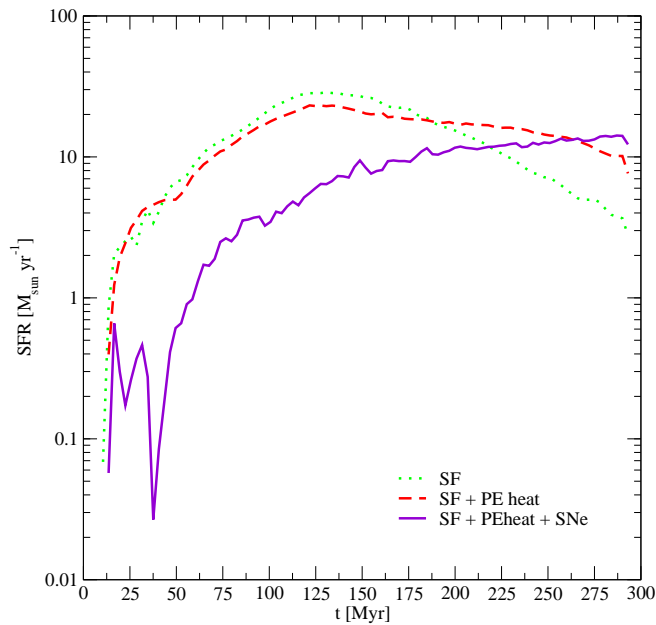}
\caption{Star formation history for simulations that include active star formation. The effect of supernova feedback strongly suppresses the star formation during the first 100\,Myr of the simulation, when the clouds are still small. 
\label{fig:starhistory}}
\end{center}
\end{figure}

The rate at which stars are being formed is plotted in Figure~\ref{fig:starhistory} as the star formation rate (SFR) history through the whole disk for the three star-forming runs. In the first 100\,Myr of the simulation, before the disk has fully fragmented, runs SFOnly and PEHeat show nearly identical SFRs, with PEHeat dipping just below SFOnly after 50\,Myr as the filament structure forms. The SNeHeat run, however, shows a strongly suppressed SFR, with sharp drops in the first 40\,Myr. During this period, the newly forming clouds are small and are efficiently disrupted by the localised feedback, shutting off their star formation after the first wave of particles form. This first epoch reduces cloud formation for all runs, as SFOnly and PEHeat runs convert the clouds into stars and SNeHeat can destroy them with thermal feedback. This is shown in the scattered start for the cloud number plotted in Figure~\ref{fig:clouds_number}. However, after 50\,Myr, the cloud population starts to grow steadily and the clouds become massive enough to survive the thermal feedback. The SFR in SNeHeat remains suppressed compared to SFOnly and PeHeat, which are almost a factor of 10 higher at their peak value at 125\,Myr. After this time, the disk is fully fragmented and both SFOnly and PEHeat begin to feel the effects of gas depletion, reducing the fuel from which to form stars. The SFR in run SFOnly drops more rapidly than for PEHeat, whose gas reduces more slowly thanks to the added support from the photoelectric heating. With a greatly reduced SFR, run SNeHeat does not feel the effects of gas depletion, and gains a steadily higher SFR throughout the simulation, overtaking SFOnly and PEHeat before 300\,Myr.

% Schmidt comparison
\begin{figure}
\begin{center}
\includegraphics[width=\columnwidth]{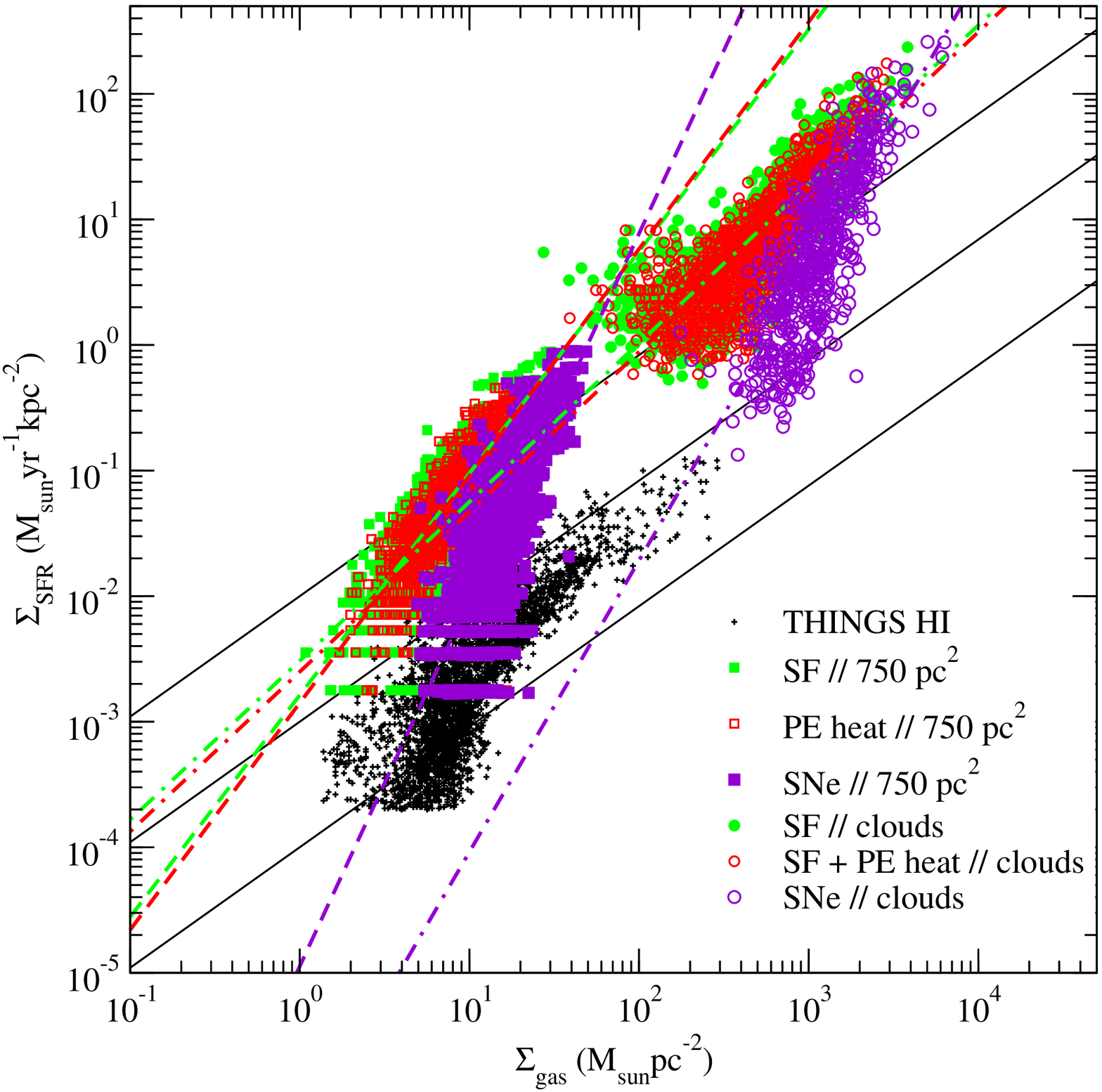}
\caption{The surface SFR, $\Sigma_{\rm sfr}$, vs. surface gas density, $\Sigma_{\rm gas}$ for the cloud populations in all three disks presented in this set of papers that include star formation at t = 200\,Myr. The surface area is taken to be in the y-z plane, i.e. as if the clouds were viewed from inside the disk. Filled green squares are for an averaged area of 750\,pc across each cloud (equivalent spatial resolution to the observations) in disk SFOnly with a best fit gradient of $\alpha_{\rm sfr} = 1.77$. Open red squares show the same averaged area around clouds in disk PEHeat with $\alpha_{\rm sfr} = 1.81$ and filled violet squares denote clouds in disk SNeHeat with $\alpha_{\rm sfr} = 2.92$ Filled green circles show the relation averaged over just the cloud's surface area in disk SFOnly, with an $\alpha_{\rm sfr} = 1.27$. Open red circles show the same in disk PEHeat and with an $\alpha_{\rm sfr} = 1.27$ and open violet circles show individual clouds in SNeHeat with $\alpha_{\rm sfr} = 2.32$. Black crosses display the observational results from the THINGS survey \citep{Bigiel2008}. The diagonal solid lines mark constant star formation efficiency, indicating the level of $\Sigma_{\rm SFR}$ required to consume 1\%, 10\% and 100\% of the gas in $10^8$\,Myr (as shown in \citet{Bigiel2008}). 
\label{fig:schmidt}}
\end{center}
\end{figure}

One of the most observed star formation relations for disk galaxies is the Kennicutt-Schmidt relation, which is plotted for all runs with star formation in Figure~\ref{fig:schmidt}. This relation links the surface star formation rate with the surface gas density and is plotted both for the individual clouds (circles) and for regions of the disk averaged over an area of 750\,pc across each cloud (squares); a region selected to match the spatial resolution of the observations in the THINGS survey of nearby galaxies \citep{Bigiel2008}. The black crosses show these observations results and the diagonal solid lines mark constant star formation efficiency. 

Without thermal feedback, the results from SFOnly and PEHeat overlap strongly, with a slight reduction in the SFR due to the photoelectric heating. Unsurprisingly, the SFR for the individual clouds lies above the averaged disk section, since this includes only gas above our star formation threshold of $100$\,cm$^{-3}$. The exponent, $\alpha_{\rm sfr}$ in the relation $\Sigma_{\rm sfr} \propto \Sigma_{\rm gas}^{\alpha_{\rm sfr}}$ is $\alpha_{\rm sfr} = 1.27$ for the clouds in both these runs and a steeper gradient corresponding to $\alpha_{\rm vir} = 1.77$ for run SFOnly and 1.81 for run PEHeat for the average disk sections. This steepening corresponds to the addition of non-star forming gas in the region and is mirrored in the observations for gas below $10$\,M$_\odot$\,pc$^{-2}$. 

While the gradient of the cloud gas for SFOnly and PEHeat shows a reasonable agreement with the higher density gas in the observations, the SFR is a factor of ten too high. This is reduced with the addition of localised thermal feedback in SNeHeat, but only for the lower density clouds, creating a steeper gradient in the cloud population with $\alpha_{\rm sfr} = 2.32$. This comes about because the smaller clouds are more strongly affected by the thermal feedback than the more massive population, as was seen in the star formation history plotted in Figure~\ref{fig:starhistory}, where the SFR showed sudden drops when the cloud population was young and small. For the biggest clouds in the simulation at surface densities above $2000$\,M$_\odot$\,pc$^{-2}$, the thermal addition makes no difference to the SFR. For the smallest clouds below $1000$\,M$_\odot$\,pc$^{-2}$, it drops it below a factor of 10. The same steepening trend is seen when the gas is averaged over the larger regions, giving an $\alpha_{\rm sfr} = 2.92$.

The fact that thermal feedback does not bring the SFR down to observed values, even when averaged over equivalent areas, suggests that our model still lacks physics that will reduce star production. This could be from another form of feedback, such as ionising winds or from a differently implemented supernovae scheme that is more efficient at dispersing the star-forming gas in larger clouds.

\section{Conclusions}

This paper presented a global simulation of a galactic disk that included star formation, photoelectric heating and localised thermal feedback at a limiting resolution of 7.8\,pc. Star forming GMCs were identified as continuous structures with gas density greater than $n_{\rm H} > 100$\,cm$^{-3}$ and their evolution was tracked through the simulation. The results were compared with three other simulations from the same series that were run with identical initial conditions but included different stellar physics, namely no star formation (run NoSF), star formation but no form of feedback (run SFOnly) and star formation with a non-localised photoelectric heating term (run PEHeat). 

Our main result suggests that while localised thermal feedback is effective at suppressing the star formation inside GMCs, the cloud itself can survive the injection of energy, which removes mass but does not cause dispersion. The suppression of star formation without cloud destruction meant that the addition of localised feedback causes the cloud properties to closely resemble those when no stellar physics is included. This implies that thermalized feedback such as supernovae may play an important role in regulating star formation, but for quiescent Milky Way-sized galaxies, the impact of thermalized feedback may be small compared to gravitational interactions and disk shear.  

 In more detail, we found:

\begin{itemize}
\item The typical (most common) cloud properties are independent of the included stellar physics. Peak mass is $6\times 10^5$\,M$_\odot$, average radius 15 - 20\,pc and a virial parameter close to the borderline value of 1.0. 99\,\% of the clouds have lifetimes less than 20\,Myr, with the difference in stellar physics only significantly affecting the lifetimes of the 1\,\% longer lived clouds. The cloud merger rate is also constant between runs, occurring every 0.2 - 0.3 of an orbital period. 

\item Without localised feedback, star formation reduces the maximum cloud mass. However, when thermal energy is injected at the star particle positions, further star formation is suppressed, resulting in a higher maximum mass that lies in-between the case without star formation and that when it is included. The same trend is observed in the maximum average cloud radius.

\item The suppression of the star formation produces clouds that are more gas-rich in the SNeHeat case than either the SFOnly or PEHeat runs. At 200\,Myr, almost 70,\% of clouds in the SNeHeat run have more than 90\,\% gas, while the majority of clouds in SFOnly and PEHeat have 10 - 20\,\% gas and 30 - 40\,\% respectively.  

\item The inclusion of photoelectric heating has the largest effect on the warm ISM. Without localised feedback, a strong filamentary structure develops that reduces the number of retrograde rotating clouds, encouraging rotation in the same sense as the galaxy. When localised feedback is added, the filaments are disrupted by the gas outflows but the mass remains in the warm phase, albeit in a less structured form. This results in a lowered fraction of retrograde clouds, but higher than when photoelectric heating is included without localised feedback. 

\item The smaller clouds' SFR is more strongly affected by the localised feedback than more massive objects. This causes a much lower SFR at the beginning of the simulation when clouds have not yet grown via mergers and a significantly lower surface SFR in low surface density gas, steepening the Kennicutt-Schmidt relation between these two quantities. 
 
\item Despite the inclusion of feedback, the star formation in the disk remains a factor of 10 higher than in observations. This suggests that either different forms of stellar feedback need to be included, such an ionising winds and radiation, or a more destructive implementation should be tried. Due to cooling losses, our addition of thermal feedback is relatively weak and should be considered a lower-bound for this type of feedback implementation. Previous work has suggested that a feedback stage prior to supernova is necessary to pre-process the star formation region. With UV ionisation, the gas density surrounding the star particle is reduced, allowing the thermal feedback to exit into the warm interstellar gas \citep{Agertz2013, Hopkins2012}. This prevents rapid cooling inside the GMC and may lead to supernovae having a stronger impact on larger scales. Different feedback implementations are left for future studies. 

\end{itemize}

\end{document}